\documentclass[aip,reprint,tight]{revtex4-1}

\usepackage[utf8]{inputenc}
\usepackage[T1]{fontenc}
\usepackage{mathptmx}
\usepackage{etoolbox}
\usepackage{amssymb}
\usepackage{color}
\usepackage{graphicx}
\usepackage{amsmath}
\usepackage{mathrsfs}
\usepackage{times}
\usepackage{subeqnarray}
\usepackage{cases}
\usepackage{bm}
\usepackage{dcolumn} 
\usepackage{import}

\begin{document}

\title{Scalable synchronization cluster in networked chaotic oscillators}

\author{Huawei Fan}
\affiliation{\mbox{School of Science, Xi'an University of Posts and Telecommunications, Xi'an 710121, China}}

\author{Yafeng Wang}
\affiliation{\mbox{Nonlinear Research Institute, Baoji University of Arts and Sciences, Baoji 721016, China}}

\author{Yao Du}
\affiliation{\mbox{School of Physics and Information Technology, Shaanxi Normal University, Xi'an 710062, China}}

\author{Haibo Qiu}
\affiliation{\mbox{School of Science, Xi'an University of Posts and Telecommunications, Xi'an 710121, China}}

\author{Xingang Wang}
\email[Corresponding author: ]{wangxg@snnu.edu.cn}
\affiliation{\mbox{School of Physics and Information Technology, Shaanxi Normal University, Xi'an 710062, China}}

\begin{abstract}
Cluster synchronization in synthetic networks of coupled chaotic oscillators is investigated. It is found that despite the asymmetric nature of the network structure, a subset of the oscillators can be synchronized as a cluster while the other oscillators remain desynchronized. Interestingly, with the increase of the coupling strength, the cluster is expanding gradually by recruiting the desynchronized oscillators {\it one by one}. This new synchronization phenomenon, which is named ``scalable synchronization cluster", is explored theoretically by the method of eigenvector-based analysis, and it is revealed that the scalability of the cluster is attributed to the unique feature of the eigenvectors of the network coupling matrix. The transient dynamics of the cluster in response to random perturbations are also studied, and it is shown that in restoring to the synchronization state, oscillators inside the cluster are stabilized in sequence, illustrating again the hierarchy of the oscillators. The findings shed new light on the collective behaviors of networked chaotic oscillators, and are helpful for the design of real-world networks where scalable synchronization clusters are concerned. 
\end{abstract}
\maketitle

\begin{quotation}
Recent progress in network science opened new avenues for the study of synchronization, among which a topic of broad interest is the emergence of synchronization clusters in complex systems of networked chaotic oscillators. In the study of cluster synchronization (CS) in networked systems, an important insight gained recently is that the formation of the clusters is crucially affected by the network symmetry, e.g., the permutation symmetries or equitable partitions. The symmetry requirement imposes a stringent restriction on the formation and properties of the synchronization clusters, e.g., the contents and sizes of the clusters are fixed and can not be modified by changing the system parameters. Yet the functionality of many natural and man-made systems requires that the synchronization clusters should be adaptive to the external environment or change with the practical demands. Typical examples of this kind include power grids (in which the set of power stations online is adjusted timely according to the consumption load), economic systems (in which the active enterprise network is adapted to the fluctuation of the economic climate), sensor networks (in which the number of active sensors is dependent of the detecting tasks), and neural networks (in which the number of neurons firing coherently is changing with the cognitive functions). A question raised in network modeling therefore is: Can we design a network in such a way that the contents and sizes of the synchronization clusters can be adjusted flexibly by changing a system parameter? Our main contribution in the present work is to demonstrate and argue that this objective can be achieved in a synthetic complex network consisting of identical chaotic oscillators. 
\end{quotation}

\section{introduction}

When an ensemble of chaotic oscillators is coupled under an intermediate strength, the oscillators can be self-organized into different clusters, with oscillators inside each cluster behaving in unison but not for oscillators from different clusters.~\cite{CS:Hansel1993,CS:Hasler1998,CS:ZY2001,SCS:Ao,CS:Scholl2013,SCS:Sorrentino,SCS:Fan,SP:Wang,CS:Laser2024} This phenomenon, known as cluster synchronization (CS), has important implications to the functionality and performance of many natural and man-made complex systems, and has received continuous interest in the fields of nonlinear science and complex systems over the past decades. In the study of CS, one of the central tasks is to infer from the network structure and the local dynamics the CS states that can emerge. For systems represented by regular networks, synchronization clusters are generated normally by the mechanism of symmetry breaking and the spatial patterns of the clusters can be inspected by the naked eye directly.~\cite{SyncConditions} In particular, the clusters are of even size and the spatial patterns of the clusters are regular. Challenges arise when dealing with complex networks, where the synchronization clusters are of diverse sizes and the cluster patterns are blurred by the random connections.~\cite{CSPattern:Chaos} To identify synchronization clusters in complex networks, one approach is leveraging the information of network symmetries.~\cite{SCS:Fu1,SCS:Nicosia,SCS:Pecora,SCS:Lin,SCS:Siddique,SCS:Wang1,EV:PK2023} Briefly, if the permutation of two nodes on the network does not change the system dynamics, the pair of nodes is regarded as symmetric, and the group of nodes permutative with each other has the potential to form a cluster. Numerical results based on computational group theory show that, despite the random connections, a complex network might possess rich symmetries, thereby supporting potentially a variety of CS states.~\cite{SCS:Pecora,SP:Ruiz-Silva} (Besides the network-symmetry-based approach, synchronization clusters in complex networks can also be analyzed by other approaches such as external equitable partition,~\cite{EEP:Cardoso,EEP:OClery,EEP:Schaub2016} simultaneous block diagonalization,~\cite{SBD:Irving,SBD:Zhang2020} and eigenvector-based analysis.~\cite{CSO:Khanra,SP:Fan})        

An intriguing CS phenomenon observed in complex networks of coupled chaotic oscillators is the coexistence of a single synchronization cluster and many desynchronized oscillators. This phenomenon, which is termed ``isolated desynchronization" in Ref.~\cite{SCS:Pecora} and ``independently synchronizable cluster" in Ref.~\cite{CSO:Cho}, is analogous to the partially coherent motions observed in a variety of realistic systems (e.g., unihemispheric sleep, neural bump states, and power grids outages), and has drawn considerable attention in nonlinear science in recent years.~\cite{SCS:Pecora,CSO:Cho,SCS:Wang2,CS:WY2023} From the perspective of network symmetry, to make a portion of the network nodes synchronized while keeping the remaining nodes desynchronized, a necessary condition is that nodes inside the cluster should be symmetric with each other.~\cite{SCS:Pecora,SCS:Siddique,EEP:Schaub2016} More specifically, nodes inside the cluster should be embraced by the same set of neighbors on the network and receive the same input signals during their time evolutions. As a consequence of this, the contents of the cluster are fixed by the network structure and can not be adjusted flexibly by tuning the system parameters such as the coupling strength. This feature of CS, however, is unfavorable and even detrimental in realistic situations where a size-changeable synchronization cluster is often required. For instance, in neuronal networks, the contents of the cluster in which the neurons are firing coherently (e.g., the bump states) are changing with the cognitive functions;~\cite{TranSyn:MS:2001,TranSyn:RC2003} in power-grid networks, it is desired that when some units are dysfunctional (desynchronized), the remaining units maintain still the synchronization state;~\cite{PowerGrid-1,PowerGrid-2} in sensor networks, all the sensors are activated during an emergency, while only a small portion of the sensors are activated in the normal situations.~\cite{SensorNet-1,SensorNet-2} An interesting question raised in network modeling is: Can we design a complex network of coupled chaotic oscillators in which the size of the synchronization cluster can be adjusted freely by changing a single system parameter?

To answer the question above, we propose in the present work a new type of network model consisting of identical chaotic oscillators, and investigate the CS behaviors emerged under different coupling strengths. The new network model is inspired by the organization of some realistic networks in nature and engineering~\cite{SensorNet-1,SN:Wasserman}, in which the network size is growing with time and the newly introduced node is connected to all the existing nodes with weighted links. We are able to demonstrate numerically and argue mathematically that the artificial network such constructed is capable of generating a scalable synchronization cluster. Specifically, amidst the desynchronization background, a subset of oscillators on the network are completely synchronized and form a cluster, while the size of the cluster can be freely adjusted by changing the uniform coupling parameter. We shall introduce the model of synthetic network in the following section. The synchronization behaviors of the synthetic network will be studied numerically in Sec. III, where the phenomenon of scalable synchronization cluster will be reported. The mechanism generating scalable synchronization cluster will be explored by the method of eigenvector-based analysis in Sec. IV. Generalization of the reported phenomenon in other network models will be studied in Sec. V. The paper is wrapped up with discussions and conclusion, which will be given in Sec. VI.

\section{Model}

The network model employed in our studies reads
\begin{equation}\label{model}
\dot{\mathbf{x}}_{i}=\mathbf{F}(\mathbf{x}_{i})+\varepsilon\sum^{N}_{j=1}w_{ij}\mathbf{H}(\mathbf{x}_{j}),
\end{equation}
where $i,j=1,2,...,N$ are the node (oscillator) indices, $\mathbf{x}_i$ denotes the state vector of node $i$, $\mathbf{F}(\mathbf{x})$ describes the nodal dynamics, $\varepsilon$ is the uniform coupling strength, and $\mathbf{H}(\mathbf{x})$ defines the coupling function. In the present work, we describe the nodal dynamics by chaotic oscillators, which are identical in dynamics but are started from different initial conditions. The coupling relationship of the oscillators is captured by the weighted matrix  $\mathbf{W}=\{w_{ij}\}$, with $w_{ij}=w_{ji}>0$ being the coupling strength between nodes $i$ and $j$. The diagonal elements of $\mathbf{W}$ are set as $w_{ii}=-\sum_{j(j\neq i)}w_{ij}$, so that $\mathbf{W}$ is a Laplacian matrix. The model described by Eq.~(\ref{model}) has been widely adopted in the literature for exploring the synchronization behaviors of coupled chaotic oscillators, and is representative of the dynamics of a variety of complex systems in nature and engineering.~\cite{SYNREV:Boccaletti,SYNREV:Arenas} 

\begin{figure}[tbp]
\begin{center}
\includegraphics[width=0.8\linewidth]{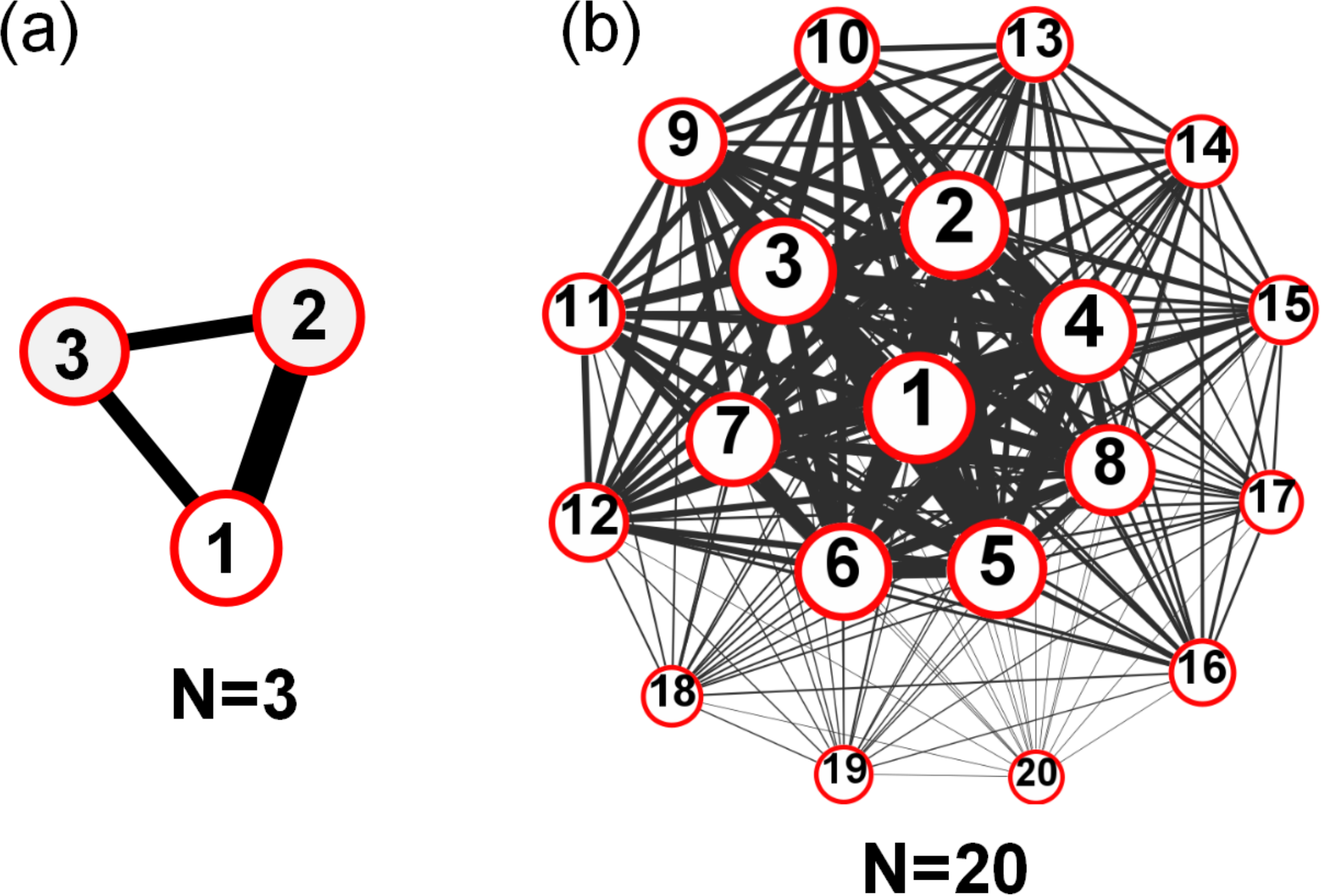}
\caption{Networks generated by the new model at the growing steps $N=3$ (a) and $N=20$ (b). The sizes of the nodes are proportional to the node capacities, and the widths of the network connections are proportional to the link weights.}
\vspace{-0.5cm}
\label{fig1}
\end{center}
\end{figure}

The key difference between the network model in the current study and the previous ones lies in the network coupling matrix $\mathbf{W}$. Whereas many network models have been proposed for describing the real-world complex systems~\cite{REV:RA,REV:MEJ}, our model is unique in that the weights (coupling strengths) of the network links are organized in a hierarchical fashion. To be specific, starting from a seeding node, the network is constructed by introducing new nodes continuously, while each new node is connected to all the existing nodes with the same weight. The network links are non-directed, and the sum of the weights associated with each new node is fixed as unity. Assume that at a time step of the network growth the existing network contains $n-1$ nodes and node $n$ is newly introduced, the coupling weight (strength) between node $n$ and each of the existing nodes is $w_{jn}=w_{nj}=1/(n-1)$, with $j=1,\ldots,n-1$. Shown in Fig.~\ref{fig1} are networks generated by the new model at two different growing steps, in which the widths of the network links are proportional to the link weights and the sizes of the nodes are proportional to the node capacities. Here, node capacity is defined as the sum of the weights calculated over all links attached to the node. In terms of the network coupling matrix, the capacity of node $i$ is $c_i=-w_{ii}$. 

Though constructed artificially, the new network model captures the key features of some real-world complex systems. For example, in social systems, when a new member joins a club, he/she will be introduced to all the existing members.~\cite{SN:Mark1973,SN:Wasserman} However, compared to the well-established relationship among the existing members, the relationship between the new member and the elder ones is relatively weak and it might take some time for them to build up the intimacy. Another example is complex sensor systems, such as environmental sensor networks and device-monitoring systems.~\cite{SensorNet-3} In these systems, to ensure an efficient collection and sharing of the acquired data, it is necessary to maintain direct communications between any pair of sensors, i.e., the sensors are globally connected. However, due to practical concerns such as energy constraints and data processing requirements, some core sensors serve as the information hubs and are strongly connected with each other, while the peripheral sensors, which are mainly used for data acquisition and upload, are weakly connected. Other examples that can be approximated by the proposed network model include military command systems, internet backbone infrastructure, and brain networks (e.g., the functional networks constructed according to the correlation of neuronal activities).

The coupling matrix of the constructed network possesses some distinct features.~\cite{chung_book} First, the network structure is asymmetric. This feature can be understood from the node capacity $c_i=1+\sum_{j>i}1/(j-1)$, which is uniquely defined for each node. One of our main findings in the present work is that despite the asymmetric nature, synchronization clusters can still emerge on the network, as will be demonstrated numerically in the following section. Second, the eigenvalues of the network coupling matrix can be explicitly given. Specifically, let $0=\lambda_{1}>\lambda_{2}\geq\ldots\lambda_k\ldots\geq\lambda_{N}$ be the eigenvalues of $\mathbf{W}$, we have 
\begin{equation}
\label{eigen}
\lambda_k=-(1+\sum_{i=N-k+1}^{N-1}\frac{1}{i}),
\end{equation}
for $k=2,\ldots,N$. Third, the eigenvectors of the network coupling matrix can also be explicitly expressed,~\cite{spielman_paper,EV:Forrow} Specifically, the elements of the eigenvector associated with $\lambda_k$ are
\begin{equation}\label{vki}
v_{k,i}=\left\{
            \begin{aligned}
           & \frac{1}{\sqrt{(N-k+1)(N-k+2)}}, & i<N-k+2,\\
            &-\frac{N-k+1}{\sqrt{(N-k+1)(N-k+2)}}, & i=N-k+2,\\
            &0, & i>N-k+2.
            \end{aligned}
            \right.
\end{equation}
Owning to these unique features, we are able to predict not only the contents of the synchronization cluster for a given coupling strength, but also the sequence of the network nodes joining in the cluster as the coupling strength increases, as will be shown theoretically in Sec. IV.

\section{Numerical results}

We start by investigating numerically the synchronization behaviors of the new network model. For demonstration purposes, we set the network size as $N=100$, and adopt the chaotic Lorenz oscillator as the nodal dynamics. The coupling function is chosen as $\mathbf{H}(\mathbf{x})=[x,0,0]^{T}$, i.e., the oscillators are coupled through their $x$ variables. The dynamics of the $i$th oscillator are described by the equations
\begin{equation}
\begin{cases}
\dot{x_{i}}=\alpha(y_{i}-x_{i})+\varepsilon\sum w_{ij}x_{j}, \\
\dot{y_{i}}=x_{i}(r-z_{i})-y_{i},\\
\dot{z_{i}}=x_{i}y_{i}-bz_{i},
\end{cases}
\label{Lorenz}
\end{equation}
with $(\alpha,r,b)=(10,28,2)$ the parameters of the Lorenz oscillators. With this set of parameters, the oscillators present the chaotic motions in the isolated form, with the largest Lyapunov exponent being about $\Lambda=0.82$.~\cite{Lorenz} In numerical simulations, the initial conditions of the oscillators are randomly chosen within the range $(-1,1)$, and Eq.~(\ref{Lorenz}) is solved by the $4$th-order Runge-Kutta algorithm with the time step $\delta t=1\times 10^{-2}$.   

\begin{figure}[tbp]
\begin{center}
\includegraphics[width=0.8\linewidth]{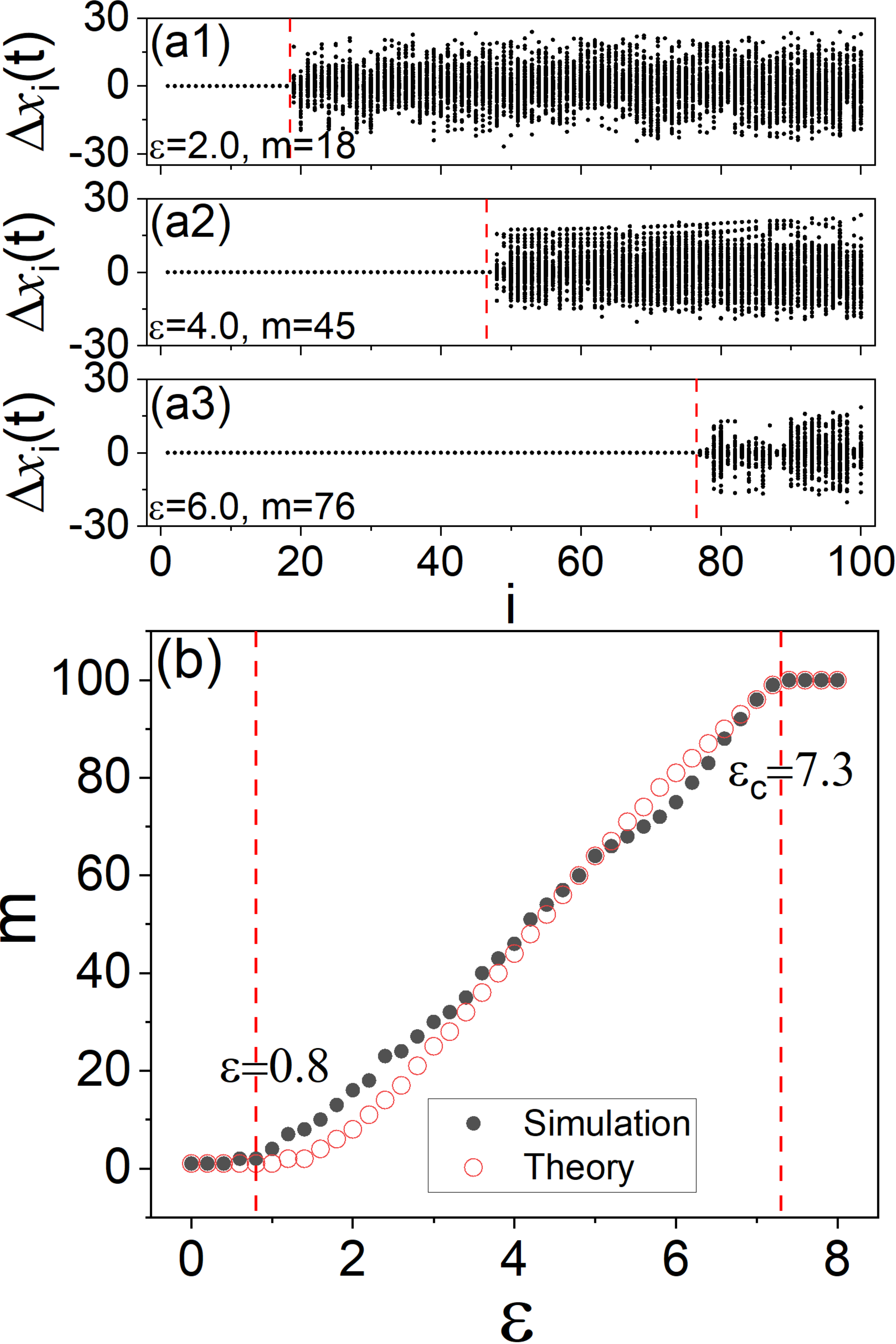}
\caption{Scalable synchronization clusters in the network containing $N=100$ chaotic Lorenz oscillators. (a) Typical CS states generated at different coupling strengths. Shown are the synchronization errors, $\Delta x_i(t)=x_i(t)-x_1(t)$, of the oscillators for $100$ successive time steps. (a1) $\varepsilon=2.0$, the cluster contains $m=18$ oscillators; (a2) $\varepsilon=4.0$, the cluster contains $m=45$ oscillators; (a3) $\varepsilon=6.0$, the cluster contains $m=76$ oscillators. (b) The variation of the cluster size, $m$, with respect to the coupling strength, $\varepsilon$. The 1st and 2nd oscillators are synchronized at $\varepsilon\approx 0.8$ (where $m$ is increased to 2). The whole network is synchronized at $\varepsilon_c\approx 7.3$. Black dots are results obtained by simulations; red circles are results predicted by the theory.}
\vspace{-0.5cm}
\label{fig2}
\end{center}
\end{figure}

The synchronization relationship between the oscillators is characterized by the error $\Delta x_i(t)=x_i(t)-x_r(t)$, with $x_r(t)$ the $x$ variable of the reference oscillator. Without the loss of generality, here we choose the $1$st oscillator (the seeding node in network construction) as the reference. If after the transient we have $\Delta x_i(t)<\Delta x_c=1\times 10^{-10}$ for the following system evolution, the $i$th oscillator is regarded as synchronized with the $1$st oscillator, and the group of oscillators with $\Delta x_i(t)<\Delta x_c$ on the network forms the synchronization cluster. Otherwise, if $\Delta x_i(t)$ is varying with time wildly after the traisent, the $i$th oscillator is regarded as not synchronized with the $1$st oscillator. Further, if $\Delta x_i(t)\neq\Delta x_j(t)$ for any pair of oscillators, both oscillators $i$ and $j$ are regarded as desynchronized. Setting the transient period as $T=1\times 10^3$, we plot in Fig.~\ref{fig2}(a) the synchronization relationship of the oscillators under several different coupling strengths. The figures are plotted by showing $100$ successive values of $\Delta x(t)$ for each oscillator. Shown in Fig.~\ref{fig2}(a1) are the results for the coupling strength $\varepsilon=2.0$. We see that $\Delta x_i(t)\approx 0$ for $i\leq18$ and $\Delta x_i(t)\neq 0$ for $i>18$. That is, oscillators of index from $1$ to $18$ are synchronized as a cluster, while the remaining oscillators are desynchronized. Shown in Fig.~\ref{fig2}(a2) are the results for $\varepsilon=4.0$. We see that in this case the synchronization cluster contains $m=45$ oscillators, with the index of the oscillators ranging from $i=1$ to $45$. It is worth noting that the synchronization cluster shown in Fig.~\ref{fig2}(a2) is expended based on the cluster shown in Fig.~\ref{fig2}(a1). Specifically, as $\varepsilon$ is increased from $2.0$ to $4.0$, $10$ more oscillators ($i=19,\ldots,45$) are escaped from the desynchronization background and join the synchronization cluster. Shown in Fig.~\ref{fig2}(a3) are the results for $\varepsilon=6.0$, in which the size of the synchronization cluster is increased further to $m=76$. 

To have a global picture of the expansion of the cluster, we plot in Fig.~\ref{fig2} the variation of the cluster size, $m$, with respect to the coupling strength, $\varepsilon$. It is seen that synchronization starts at about $\varepsilon=0.8$, where oscillators $1$ and $2$ are synchronized and form the core of the cluster. After that, with the increase of $\varepsilon$, the cluster is expanded gradually by recruiting oscillators from the desynchronization background one by one in the ascending order of the oscillator index. Finally, at about $\varepsilon_c=7.3$, all oscillators are within the cluster and the network is globally synchronized. As the size of the cluster can be adjusted freely by tuning the coupling strength, the new type of synchronization behavior is termed scalable synchronization cluster in our present work.

\begin{figure*}[tbp]
\begin{center}
\includegraphics[width=0.67\linewidth]{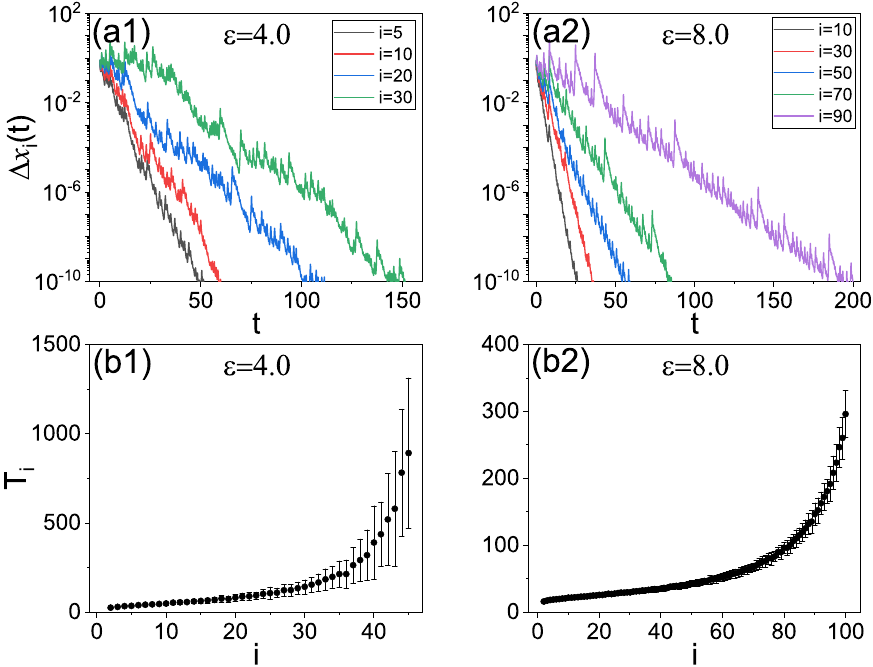}
\caption{Transient dynamics of the synchronization cluster in response to random perturbations. Shown are the time evolution of the synchronization error $\Delta x_i(t)$ for typical oscillators inside the cluster (upper panels) and the distribution of the transient time, $T_i$ (lower panels). (a1,b1) are results for the coupling strength $\varepsilon=4.0$, in which the synchronization cluster contains $m=45$ oscillators. (a2,b2) are the results for $\varepsilon=8.0$, in which the network is globally synchronized. Numerical results in (b1) and (b2) are averaged over $100$ perturbation realizations. Vertical bars denote the variances.}
\vspace{-0.5cm}
\label{fig3}
\end{center}
\end{figure*}

The hierarchy of the oscillators is reflected in not only the expansion of the cluster, but also the transient dynamics of the cluster in restoring to the synchronization state. To demonstrate, we take the CS state shown in Fig.~\ref{fig2}(a2) as the example ($\varepsilon=4.0$, $m=45$), and introduce random perturbations to all oscillators and check the responses of the system dynamics. Here, the (independent and identically distributed) perturbations are chosen randomly within the range $(-1,1)$, and are added onto all variables of each oscillator at only the beginning. It is found that while the network is restored to the original CS state eventually, the transient dynamics of the oscillators are of clear difference. Shown in Fig.~\ref{fig3}(a1) are the transient behaviors of some typical oscillators in the cluster ($i=(5,10,20,30,40)$). Again, we characterize the responses of the oscillators by the synchronization error $\Delta x_i(t)$. The results in Fig.~\ref{fig3}(a1) suggest that the oscillators are stabilized in sequence in restoring to the CS state. Defining the transient time of oscillator $i$ as the moment when $\Delta x_i(t)<1\times 10^{-10}$, we plot in Fig.~\ref{fig3}(b1) the transient time $T_i$ for all oscillators inside the cluster ($i=1,\ldots,45$). Indeed, we see that the oscillators are stabilized by the order of the oscillator index, signifying the hierarchy of the oscillators in the temporal domain. Similar results are also observed for the global synchronization state, as depicted in Fig.~\ref{fig3}(a2) and (b2). 

\section{Theoretical analysis}

How could synchronization clusters be generated in asymmetric complex networks and why the size of the cluster can be adjusted freely by tuning the coupling strength? Intrigued by these questions, we proceed to conduct a theoretical analysis on the generating mechanism of the CS states. 

\begin{figure}[tbp]
\begin{center}
\includegraphics[width=0.8\linewidth]{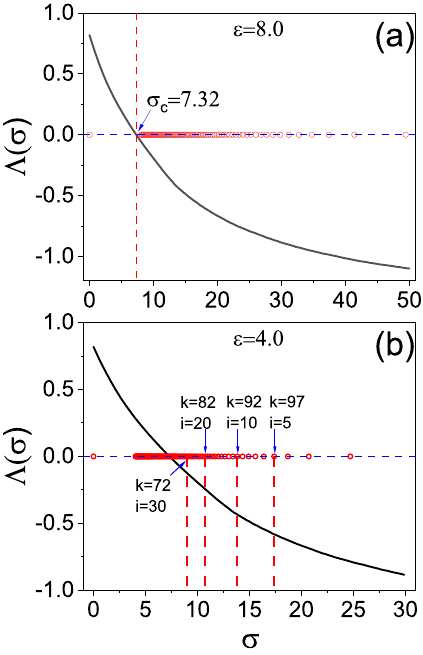}
\caption{(a) The MSF of chaotic Lorenz oscillators coupled through the $x$-variable. $\Lambda<0$ for $\sigma>\sigma_c\approx 7.32$. Shown are also the distribution of the modes for the coupling strength $\varepsilon=8.0$. As $\varepsilon$ decreases, modes in the stable domain will be shifted into the unstable domain in sequence. (b) The distribution of the modes for $\varepsilon=4.0$. $\tilde{m}=55$ modes are in the unstable domain and $\hat{m}=45$ modes are in the stable domain. Shown are also the stability of four typical modes (oscillators) in the stable domain (cluster), $k=(72,82,92,97)$.}
\vspace{-0.5cm}
\label{msf-lorenz}
\end{center}
\end{figure}

As the chaotic oscillators are linearly coupled and $\mathbf{W}$ is a Laplacian matrix, the global synchronization state is a solution for Eq.~(\ref{model}). Denoting $\mathbf{s}(t)$ as the manifold of the global synchronization state, i.e., $\mathbf{s}(t)=\mathbf{x}_{1}(t)=\mathbf{x}_{2}(t)=\ldots=\mathbf{x}_{N}(t)$, the starting point of our analysis is to evaluate the stability of the global synchronization state in the presence of small perturbations. Let $\delta \mathbf{x}_{i}=\mathbf{x}_{i}-\mathbf{s}$ be the infinitesimal perturbation added on oscillator $i$ in the global synchronization state, the time evolution of $\delta \mathbf{x}_{i}$ is governed by the variational equations
\begin{equation}\label{variational-eq}
\delta\dot{\mathbf{x}}_{i}=\mathbf{DF}(\mathbf{s})\delta\mathbf{x}_{i}+\varepsilon\sum^{N}_{j=1}w_{ij}\mathbf{DH}(\mathbf{s})\delta\mathbf{x}_{j},
\end{equation}
where $\mathbf{DF}(\mathbf{s})$ and $\mathbf{DH}(\mathbf{s})$ are the Jacobian matrices evaluated on the synchronization manifold. Transforming Eq.~(\ref{variational-eq}) into the space spanned by the eigenvectors of $\mathbf{W}$, we have
\begin{equation}\label{decoupled-eq}
\delta\mathbf{\dot{y}}_{k}=[\mathbf{DF}(\mathbf{s})+\varepsilon\lambda_{k}\mathbf{DH}(\mathbf{s})]\delta\mathbf{y}_{k},
\end{equation}
with $k=1,\ldots,N$. In Eq.~(\ref{decoupled-eq}), $\Delta\mathbf{Y}=[\delta \mathbf{y}_{1}, \delta \mathbf{y}_{2},..., \delta \mathbf{y}_{N}]^{T}=\mathbf{V}^{-1}[\delta \mathbf{x}_{1}, \delta \mathbf{x}_{2},..., \delta \mathbf{x}_{N}]^{T}$ are the perturbation modes in the new space, with $\mathbf{V}$ being the transformation matrix composed by the eigenvectors of $\mathbf{W}$ and $0=\lambda_{1}>\lambda_{2}\geq\ldots\geq\lambda_{N}$ being the eigenvalues of $\mathbf{W}$. Among the $N$ modes, the one associated with $\lambda_1$ describes the motion parallel to the synchronization manifold, and the modes associated with $\lambda_{2,\ldots,N}$ describe the motions transverse to the synchronization manifold. Following the formalism of master stability function (MSF),~\cite{MSF-1,MSF-2,MSF-3} we introduce the generic coupling strength $\sigma\equiv-\varepsilon\lambda$ and rewrite Eq.~(\ref{decoupled-eq}) as
\begin{equation}\label{msf}
\delta\mathbf{\dot{y}}_{k}=[\mathbf{DF}(\mathbf{s})-\sigma_k\mathbf{DH}(\mathbf{s})]\delta\mathbf{y}_{k}.
\end{equation}
For the global synchronization state to be stable, the necessary condition is that all the perturbation modes transverse to the synchronization manifold should be damping with time. That is, the largest conditional Lyapunov exponent, $\Lambda_k$, as calculated from Eq.~(\ref{msf}) should be negative for all the transverse modes ($k=2,\ldots,N$). As Eq.~(\ref{msf}) applies to all the perturbation modes, it is named the master equation of the perturbation dynamics.~\cite{MSF-1} By solving the master equation numerically, we can obtain the variation of $\Lambda$ with respect to $\sigma$, which gives the MSF curve. In the MSF curve, the regions with $\Lambda<0$ constitute the stable domain in the parameter space, and the regions with $\Lambda>0$ constitute the unstable domain. Depending on the nodal dynamics and the coupling function, the stable domain may have different forms, e.g., bounded or unbounded~\cite{MSF-3}. Presented in Fig.~\ref{msf-lorenz}(a) is the MSF curve for the chaotic Lorenz oscillators coupled through the $x$-variable, we see that $\Lambda<0$ when $\sigma>\sigma_c\approx 7.32$, i.e., the stable domain is unbounded. Therefore, the global synchronization state is stable when $\varepsilon>\varepsilon_c\equiv\varepsilon_2=-\sigma_c/\lambda_2$, with $\lambda_2$ the 2nd largest eigenvalue of $\mathbf{W}$. The above analysis is identical to the standard MSF formalism,~\cite{MSF-1,MSF-2,MSF-3}, which focuses on only the critical coupling strength for achieving global synchronization. In what follows, we are going to demonstrate that by incorporating the information of the eigenvectors of $\mathbf{W}$, this formalism can be generalized and utilized to analyze the emergence of scalable synchronization clusters.

For the network model we have studied numerically above ($N=100$), the 2nd largest eigenvalue is $\lambda_2=-N/(N-1)\approx -1.01$. The critical coupling for global synchronization therefore is estimated to be $\varepsilon_c=-\sigma_c/\lambda_2\approx 7.25$. This estimation is consistent with the numerical result shown in Fig.~\ref{fig2}(b), where the network is globally synchronized when $\varepsilon>\varepsilon_c\approx 7.3$. The strategy we adopt for exploring the CS states emerged at the intermediate coupling strengths is network desynchronization, starting from the global synchronization state generated at a strong coupling $\varepsilon>\varepsilon_c$. To be specific, we shift the transverse modes out of the stable domain in the MSF curve one by one through decreasing $\varepsilon$, and check what happens to the system dynamics when each mode becomes unstable. A schematic of the desynchronization strategy is shown in Fig.~\ref{msf-lorenz}(a), where the coupling strength is chosen as $\varepsilon=8.0$ and all the transverse modes are within the stable domain. As $\varepsilon$ decreases, the $N-1$ transverse modes will be shifted to the left globally, with the mode associated with $\lambda_2$ being the first to become unstable (at $\varepsilon_c$). Since the mode associated with $\lambda_1=0$ is always unstable, we thus have for this situation two unstable modes, $\delta\mathbf{y}_1$ and $\delta\mathbf{y}_2$. Transforming the unstable modes, $\delta\mathbf{y}_1(t)$ and $\delta\mathbf{y}_2(t)$, into the node space, we have
\begin{equation}\label{v2}
\delta\mathbf{x}_{i}(t)=v_{1,i}\delta\mathbf{y}_{1}(t)+v_{2,i}\delta\mathbf{y}_{2}(t),
\end{equation}
where $\mathbf{v}_1=\{v_{1,i}\}_{i=1,\ldots,N}$ is the eigenvector associated with $\lambda_1$, and $\mathbf{v}_2=\{v_{2,i}\}_{i=1,\ldots,N}$ is the eigenvector associated with $\lambda_2$. As $\lambda_1=0$, we have $v_{1,i}=1/\sqrt{N}$. As such, the first term on the right-hand-side (RHS) of Eq.~(\ref{v2}) is identical for all oscillators in the network, and the perturbation that oscillator $i$ is away from the global synchronization state is solely determined by the second term on the RHS of Eq.~(\ref{v2}). With this property, Eq.~(\ref{v2}) can be rewritten as
\begin{equation}\label{v2-2}
\delta\mathbf{x}_{i}(t)=c(t)+v_{2,i}\delta\mathbf{y}_{2}(t),
\end{equation} 
where $c(t)$ denotes the uniform background and is oscillator-independent. Thus, the stability of the oscillators can be evaluated individually by checking the elements $v_{2,i}$: the larger is $|v_{2,i}|$, the far is oscillator $i$ diverged from the global synchronization state. In particular, if $v_{2,i}=v_{2,j}$, the instant states of oscillators $i$ and $j$ will be identical throughout the process of system evolution, i.e., they are completely synchronized. The set of oscillators with the same eigenvector element thus forms a synchronization cluster. As depicted in Eq.~(\ref{vki}), when the mode of $\lambda_2$ becomes unstable, we have $v_{2,i}=1/\sqrt{N(N-1)}$ for $i=1,\ldots,N-1$ and $v_{2,N}=-(N-1)/\sqrt{N(N-1)}$. The elements of $\mathbf{v}_2$ thus indicate that in this case oscillators from $1$ to $N-1$ will be synchronized and form a giant cluster, while the $N$th oscillator is desynchronized. This special CS state is observable when the coupling strength is within the interval $\varepsilon\in (\varepsilon_c,\varepsilon_3)$, with $\varepsilon_3=-\sigma_c/\lambda_3\approx 6.9$. 

As $\varepsilon$ crosses $\varepsilon_3$ from above, the mode associated with $\lambda_3$ will be also unstable. In this case, we have in total $3$ unstable modes, $\delta\mathbf{y}_1$, $\delta\mathbf{y}_2$ and $\delta\mathbf{y}_3$. Transforming them into the node space, we have the synchronization errors
\begin{equation}\label{v2-3}
\delta\mathbf{x}_{i}(t)=c(t)+v_{2,i}\delta\mathbf{y}_{2}(t)+v_{3,i}\delta\mathbf{y}_{3}(t).
\end{equation} 
Still, $c(t)$ represents the synchronization background and is identical for all oscillators in the network. Equation ~(\ref{v2-3}) suggests that if $v_{2,i}=v_{2,j}$ and $v_{3,i}=v_{3,j}$, the relation $\delta\mathbf{x}_i(t)=\delta\mathbf{x}_j$(t) will be held throughout the process of system evolution, i.e., oscillators $i$ and $j$ are completely synchronized during the process of network evolution. For the eigenvector $\mathbf{v}_3$, we have from Eq.~(\ref{vki}) that $v_{3,i}=1/\sqrt{(N-1)(N-2)}$ for $i=1,\ldots,N-2$, $v_{3,N-1}=-(N-2)/\sqrt{(N-1)(N-2)}$, and $v_{3,N}=0$. Since oscillator $N$ is already desynchronized, the network now comprises one giant cluster (oscillators from $i=1$ to $N-2$) and two desynchronized oscillators ($i=N-1$ and $N$). Therefore, as $\varepsilon$ decreases from $\varepsilon_3$, oscillator $N-1$ will be also desynchronized from the cluster, making the cluster size decreased by one (now the cluster contains $m=N-2$ oscillators). This scenario of oscillator desynchronization continues until only one oscillator is left, which is just the seeding node ($i=1$) in constructing the network. For the general case when $\tilde{m}$ modes are unstable, the synchronization errors of the oscillators are
\begin{equation}\label{vk}
\delta\mathbf{x}_{i}(t)=c(t)+\sum_{k=2}^{\tilde{m}}v_{k,i}\delta\mathbf{y}_{k}(t).
\end{equation}
In this case, the synchronization cluster contains $m=N-\tilde{m}$+1 oscillators, with the oscillator index ranging from $i=1$ to $m$.
As the eigenvalues can be explicitly obtained [see Eq.~(\ref{eigen}], we are able to predict the range over which the synchronization cluster containing $m$ oscillators is stable 
\begin{equation}
\label{range}
\varepsilon\in(\varepsilon_{\tilde{m}+1},\varepsilon_{\tilde{m}})=(-\sigma_c/\lambda_{N-m+2},-\sigma_c/\lambda_{N-m+1}).
\end{equation}  

The above theoretical predictions are in good agreement with the numerical results reported in the previous section. For example, by the coupling strength $\varepsilon=4.0$ studied in Fig.~\ref{fig2}(a2), we plot in Fig.~\ref{msf-lorenz}(b) the distribution of the modes in the parameter space, which shows that there are in total $\tilde{m}=55$ unstable modes. According to our above analysis, the synchronization cluster should contain $m=N-\tilde{m}+1=45$ oscillators ($i=1,\ldots,45$), which agrees with the numerical results shown in Fig.~\ref{fig2}(a2). By Eq.~(\ref{range}), we plot in Fig.~\ref{fig2}(b) the variation of the cluster size, $m$, with respect to the coupling strength, $\varepsilon$ (the red circles). It is seen that the critical couplings predicted by the theory are in good agreement with the numerical results over the whole parameter space.

Owing to the unique feature of the eigenvectors, the transient dynamics of the CS state in response to perturbations can also be analyzed. Different from the above analysis which focuses on the stability of the unstable modes, this time we are interested in the transient dynamics of the stable modes. Considering still the general case when $\tilde{m}$ modes are unstable. The number of stable modes is $\hat{m}=m-1$, with $m$ the size of the synchronization cluster. Focusing on only oscillators inside the cluster, $j=1,\ldots,m$, the evolution of the perturbation error of oscillator $j$ is governed by the equation
\begin{equation}\label{transient-1}
\delta\mathbf{x}_{j}(t)=\sum_{k=1}^{\tilde{m}}v_{k,j}\delta\mathbf{y}_{k}(t)+\sum_{k=\tilde{m}+1}^{N}v_{k,j}\delta\mathbf{y}_{k}(t).
\end{equation}
Please note that the first term on the RHS of Eq.~(\ref{transient-1}) is contributed by the unstable modes, which is identical for oscillators in the cluster and therefore can be treated as the uniform perturbation background [similar as $c(t)$ in Eq.~(\ref{vk})]. Defining $\Delta \mathbf{x}^c_{j}(t)\equiv \delta\mathbf{x}_{j}(t)-\sum_{k=1}^{\tilde{m}}v_{k,j}\delta\mathbf{y}_{k}(t)$ as the cluster-based synchronization error, Eq.~(\ref{transient-1}) can be simplified as
\begin{equation}\label{transient-2}
\Delta\mathbf{x}^c_{j}(t)=\sum_{k=\tilde{m}+1}^{N}v_{k,j}\delta\mathbf{y}_{k}(t).
\end{equation}
For the case of infinitesimal random perturbations, we have $\delta\mathbf{y}_{k}(t)\approx \delta\mathbf{y}_{0}\exp{\Lambda_k t}$, with $\delta\mathbf{y}_{0}$ the initial perturbation and $\Lambda_k$ the largest Lyapunov exponent of the $k$th mode as caculated from Eq.~(\ref{msf}). Regarding this, Eq.~(\ref{transient-2}) can be rewritten as 
\begin{equation}\label{transient-3}
\Delta\mathbf{x}^c_{j}(t)=(\sum_{k=\tilde{m}+1}^{N}v_{k,j}e^{\Lambda_k t})\delta\mathbf{y}_{0},
\end{equation}
which describes how the synchronization errors between oscillators inside the cluster are evolving with time.

As depicted by the MSF curve shown in Fig.~\ref{msf-lorenz}, the value of $\Lambda_k$ is decreased with $k$ monotonically. That is, the stable modes will be damped to zero in sequence, starting from the mode $k=N$ and ending with the mode $k=\tilde{m}+1=N-m+2$. When the $N$th mode is damped to zero, the contribution of this mode to $\Delta\mathbf{x}^c_{j}$ on the RHS of Eq.~(\ref{transient-3}) will be disappeared. As $v_{k,1}=v_{k,2}$ for $k=\tilde{m}+1$ to $N-1$ [as can be seen from Eq.~(\ref{vki})], the relation $\Delta\mathbf{x}^c_{1}(t)=\Delta\mathbf{x}^c_{2}(t)$ is held. That is, oscillators $1$ and $2$ are synchronized first in restoring to the CS state, with the synchronization speed (the transient time) being determined by $\Lambda_N$. Similarly, when the mode $N-1$ is damped to zero, the term of $k=N-1$ on the RHS of Eq.~(\ref{transient-3}) will be disappeared. As in this case we have $v_{k,1}=v_{k,2}=v_{k,3}$ for $k=\tilde{m}+1$ to $N-2$, the relation $\Delta\mathbf{x}^c_{1}(t)=\Delta\mathbf{x}^c_{2}(t)=\Delta\mathbf{x}^c_{3}(t)$ is held. That is, the $3$rd oscillator is restored to the synchronization state after oscillators $1$ and $2$. Still, the transient time for the $3$rd oscillator to be restored to the synchronization state is determined by $\Lambda_{N-1}$. This process of synchronization restoration continues until the $m$th oscillator is synchronized, which completes the transient process. 

\begin{figure}[tbp]
\begin{center}
\includegraphics[width=0.8\linewidth]{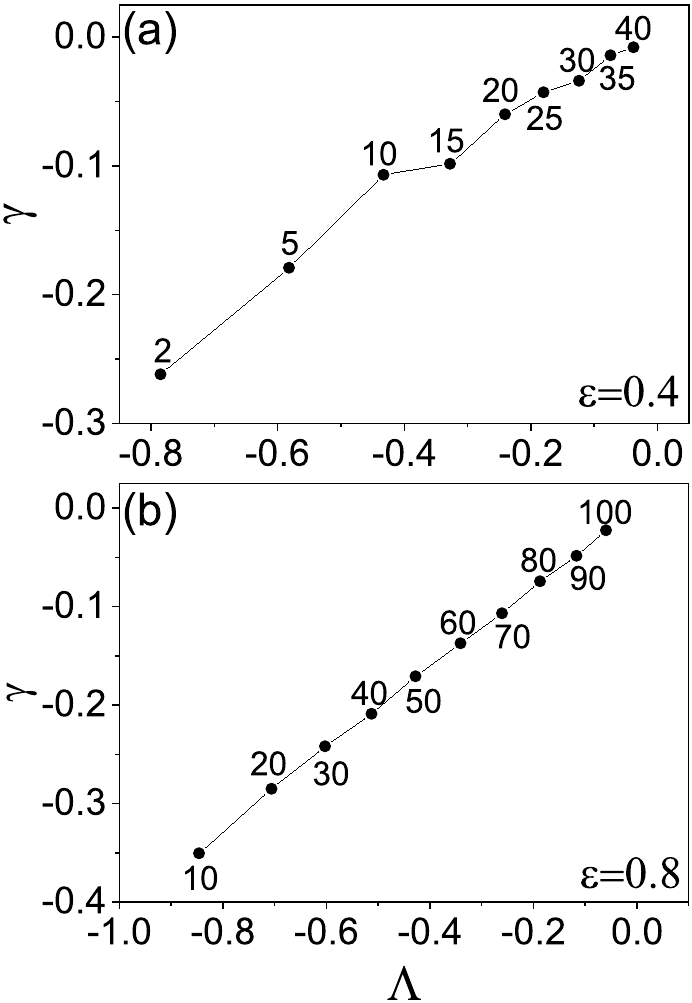}
\caption{Transient dynamics of oscillators in the cluster in restoring synchronization. $\gamma$ is the fitted exponent characterizing the synchronization speed of the oscillator; $\Lambda$ is the largest Lyapunov exponent characterizing the stability of the oscillator. (a) $\varepsilon=4.0$. (b) $\varepsilon=8.0$. Numbers in (a) and (b) denote the oscillator indices.}
\vspace{-0.5cm}
\label{transient}
\end{center}
\end{figure}

The above theoretical predictions are validated by simulations. Shown in Fig.~\ref{msf-lorenz}(b) is the distribution of the perturbation modes along the MSF curve for the coupling strength $\varepsilon=4.0$. We see that the modes from $k=56$ to $100$ are located in the stable domain. In addition, the stability of the modes, as characterized by the Lyapunov exponent $\Lambda_k$, is increased with mode index $k$. In particular, we have $\Lambda_k\approx -0.6$, $-0.45$, $-0.24$, and $-0.12$ for modes $k=97$ (the $5$th oscillator), $92$ (the $10$th ocillator), $82$ (the $20$th oscillator), and $72$ (the $30$th oscillator), respectively. Our theory thus predicts that the modes should be damped in sequence, with the damping exponents, $\gamma_k$, being correlated with the corresponding Lyapunov exponents, $\Lambda_k$. The predictions are consisent with the numerical results shown in Fig.~\ref{fig3}(a1), where the fitted damping exponents for the oscillators $5$, $10$, $20$, and $30$ are $\gamma=-0.17$, $-0.1$, $-0.06$, and $-0.04$, respectively. To validate the theoretical predictions further, we plot in Fig.~\ref{transient}(a) the relationship between $\gamma$ and $\Lambda$ for some typical oscillators in the synchronization cluster. We see in the figure that $\gamma$ is increased linearly with $\Lambda$, signifying the strong correlation between the two exponents. Similar results are also for the global synchronization state generated by the coupling strength $\varepsilon=8.0$, as depicted in Fig.~\ref{transient}(b).  

\section{Generalization}

Our above analysis indicates that the key to generating scalable synchronization clusters lies in the well-organized elements of the eigenvectors, which in turn is attributed to the hierarchical network structure. With this understanding, it is natural to expect that scalable synchronization cluster is a typical phenomenon for the proposed network model. To check it out, we proceed to investigate the dependence of the CS states on the network details, including the network size, the weight perturbations, and the oscillator dynamics.

\begin{figure}[tbp]
\begin{center}
\includegraphics[width=0.8\linewidth]{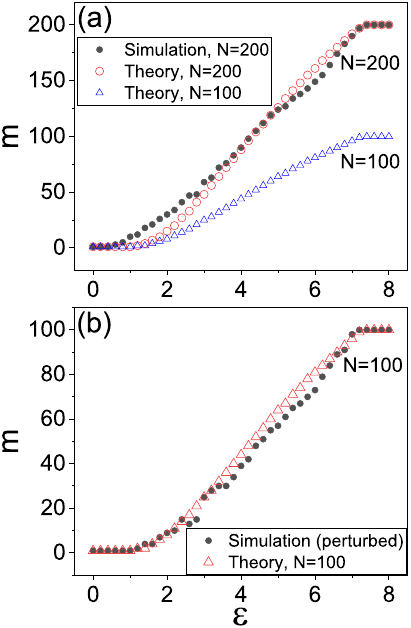}
\caption{(a) Scalable synchronization clusters in the network containing $N=200$ chaotic Lorenz oscillators. Shown is the variation of the size of the synchronization cluster, $m$, with respect to the coupling strength, $\varepsilon$. The results for $N=100$ are represented by blue triangles. (b) $m$ versus $\varepsilon$ for the network of size $N=100$ and perturbed coupling strengths (the black dots). Results predicted by the theory are represented by red triangles.}
\vspace{-0.5cm}
\label{n200}
\end{center}
\end{figure}

We investigate first the impact of the network size on the phenomenon of scalable synchronization clusters. In exploring the synchronization behaviors of complex networks, a prevailing observation is that as the network size grows, the tendency toward global synchronization will be diminished.~\cite{SYNREV:Boccaletti,SYNREV:Arenas} Different from this observation, here we find that for the proposed network model, the increase of the network size is in favor of cluster synchronization. Assume that the network contains $N$ oscillators and the mission is to generate a synchronization cluster of size $m$ (the number of unstable modes is $\tilde{m}=N-m+1$), according to Eqs.~(\ref{eigen}) and (\ref{range}), the critical coupling required for generating the cluster is
\begin{equation}\label{sizeeffect}
\bar{\varepsilon}_m=\varepsilon_{\tilde{m}}=-\sigma_c/\lambda_{\tilde{m}+1}=\sigma_c/(1+\sum_{i=m-1}^{N-1}\frac{1}{i}).
\end{equation} 
Clearly, $\bar{\varepsilon}_m$ is decreased monotonically by increasing $N$. To verify this prediction, we increase the network size to $N=200$ and, based on the results of numerical simulations, plot in Fig.~\ref{n200}(a) the variation of $m$ with respect to $\varepsilon$ (the black dots). We see that the synchronization cluster is still scalable and, compared to the results of the smaller-size network ($N=100$, the blue triangles), the critical coupling for generating a cluster of the same size is decreased. 

We next investigate the robustness of scalable synchronization clusters to weight perturbations. In constructing the network model, the new node is connected to all the existing nodes with equal weight, i.e., $w_{ij}=w_{ji}=1/n$, with $i=n+1$ denoting the new node and $j=1,\ldots,n$ being the existing nodes. This setting is impractical for real-world systems, as mismatches and perturbations are unavoidable for the weights of the network links. A question of interest therefore is whether scalable synchronization clusters can still be observed in the presence of weight perturbations (mismatches). To investigate, we adopt the network model studied in Sec. III ($N=100$) and replace $w_{ij}$ with $\xi_{ij}w_{ij}$ for each element in the network coupling matrix, with $\xi$ being randomly chosen within the range $(-1.1,1.1)$. Due to the presence of random perturbations, the elements of the eigenvectors are non-identical. Specifically, for the eigenvector $\mathbf{v}_k$, small perturbations will be introduced to the elements $v_{k,i}$ for $i<N-k+2$ [please see Eq.~(\ref{vki})]. As such, when mode $k$ becomes unstable, the synchronization errors of the oscillators with index from $i=1$ to $N-k+2$ will be slightly different from each other, making complete synchronization impossible. However, Eq.~(\ref{vk}) implies that given the perturbations are small ($\delta v_{k,i}\ll v_{k,i}$), oscillators inside the cluster can still be well synchronized. This prediction is verified by simulations, as shown in Fig.~\ref{n200}(b) (the black dots). In plotting Fig.~\ref{n200}(b), two oscillators are regarded as synchronized if after a transient ($T=200$) the synchronization error between them, $\Delta x_{ij}=|x_i-x_j|$, is smaller than the threshold $\Delta x_c=0.1$. We see that the results of the perturbed network are similar to those of the unperturbed network, signifying the robustness of the synchronization clusters to weight perturbations.       

\begin{figure}[tbp]
\begin{center}
\includegraphics[width=0.8\linewidth]{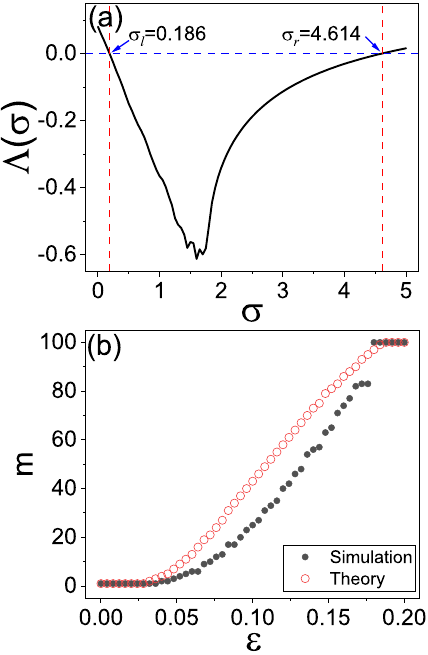}
\caption{Scalable synchronization in the network of $N=100$ identical chaotic R\"{o}ssler oscillators. (a) The MSF curve. (b) The variation of the size of the synchronization cluster, $m$, with respect to the coupling strength, $\varepsilon$. Black dots are results obtained by numerical simulations. Red circles are results predicted by the theory.}
\vspace{-0.5cm}
\label{rossler}
\end{center}
\end{figure}

We finally study the generality of the reported phenomenon by considering other types of nodal dynamics, including the chaotic R\"{o}ssler oscillator and the chaotic Hindmarsh-Rose (HR) oscillators. Different from the case studied in Sec. III (where the stable domain of the MSF curve is unbounded), here we are interested in situations where the stable domain of the MSF curve is bounded. We present first the results for chaotic R\"{o}ssler oscillators. The dynamics of the $i$th R\"{o}ssler oscillator in the network is governed by the equations
\begin{equation}
\begin{cases}
\dot{x_{i}}=y_{i}-z_{i}+\varepsilon\sum w_{ij}x_{j}, \\
\dot{y_{i}}=x_{i}+0.2y_{i},\\
\dot{z_{i}}=0.2+(x_{i}-9)z_{i}.
\end{cases}
\end{equation}
In the isolated form, the oscillators present chaotic motions, with the largest Lyapunov exponent being about $\Lambda=0.08$.~\cite{Rossler} For the adopted coupling function $\mathbf{H}(\mathbf{x})=[x,~0,~0]^{T}$, the stable domain of the MSF curve is bounded, i.e., $\Lambda<0$ for $\sigma_l<\sigma<\sigma_r$, with $\sigma_l\approx 0.186$ and $\sigma_r\approx 4.614$ denoting, respectively, the left and right boundaries of the stable domain [as shown in Fig.~\ref{rossler}(a)].~\cite{MSF-3} Still, we set the network size as $N=100$. For the case of bounded stable domain, the necessary conditions for achieving global synchronization are $\Lambda_2<0$ and $\Lambda_N<0$, which give the range over which the global synchronization state is stable, $\varepsilon\in (\varepsilon_2,\varepsilon_N)$. Here, $\varepsilon_2=-\sigma_l/\lambda_2$ and $\varepsilon_N=-\sigma_r/\lambda_N$ are the critical couplings associated with the extreme modes $\lambda_2$ and $\lambda_N$, respectively. As $\lambda_2\approx -1.01$ and $\lambda_N\approx -6.177$ for $N=100$, we have $\varepsilon_2\approx 0.184$ and $\varepsilon_N\approx 0.747$. By numerical simulations, we increase $\varepsilon$ from $0$ to $0.2$ and plot in Fig.~\ref{rossler}(b) the variation of the size of the synchronization cluster. Again, a scalable synchronization cluster is observed. By Eqs.~(\ref{eigen}) and (\ref{range}), we plot in Fig.~\ref{rossler}(b) also the theoretical results predicted by the theory. We see that the theoretical predictions are in good agreement with the numerical results. 

For MSF curves with bounded stable domains, an intriguing phenomenon is the desynchronization of the oscillators when the coupling strength is very large.~\cite{SWBSI,SYNREV:Boccaletti} Specifically, as $\varepsilon$ increases from $\varepsilon_N$, the modes will become unstable in sequence, with the mode of $\lambda_N$ being the first one and the mode of $\lambda_2$ being the last one. In this case, the critical couplings of the modes can still be estimated from Eq.~(\ref{range}), yet the oscillators will be desynchronized from the cluster in the reverse order. To be specific, as $\varepsilon$ increases from $\varepsilon_N$, it is the seeding oscillators $i=1$ and $2$ that is firstly desynchronized from the global synchronization state, and then the 3rd oscillator and so on. The desynchronization sequence can also be inferred from the hierarchical structure of the eigenvectors [as depicted in Eq.~(\ref{vki})]. Considering the situation when $\tilde{m}$ modes are unstable, the synchronization errors of the oscillators are
\begin{equation}\label{vk2}
\delta\mathbf{x}_{i}(t)=c(t)+\sum_{k=N-\tilde{m}+1}^{N}v_{k,i}\delta\mathbf{y}_{k}(t).
\end{equation}    
The same as in Eq.~(\ref{vk}), $c(t)$ stands still for the synchronization background and is identical for all the oscillators. When only mode $N$ is unstable (which occurs when $\varepsilon>-\sigma_r/\lambda_N\approx 0.747$), Eq.~(\ref{vk2}) is simplified as $\delta\mathbf{x}_{i}(t)=c(t)+v_{N,i}\delta \mathbf{y}_{N}(t)$. As $v_{N,1}=-v_{N,2}=1/\sqrt{2}$ and $v_{N,i}=0$ for $i=3,\ldots,N$, we have $\delta\mathbf{x}_{i}(t)=\delta\mathbf{x}_{j}(t)$ for $i,j>2$ and $\delta\mathbf{x}_{1}(t)\neq \delta\mathbf{x}_{2}(t) \neq 0$. That is, oscillators from $i=3$ to $N$ are synchronized as a cluster, while oscillators $1$ and $2$ are desynchronized. Similarly, when mode $N-1$ becomes unstable, we have $\delta\mathbf{x}_{i}(t)=\delta\mathbf{x}_{j}(t)$ for $i,j=4,\ldots,N$ and $\delta\mathbf{x}_{1}(t)\neq \delta\mathbf{x}_{2}(t) \neq \delta\mathbf{x}_{3}(t)\neq 0$. That is, the $3$rd oscillator is desynchronized from the cluster. This process of oscillator desynchronization continues till mode $2$ becomes unstable, which happens when $\varepsilon>-\sigma_r/\lambda_2\approx 4.57$.

\begin{figure}[tbp]
\begin{center}
\includegraphics[width=0.8\linewidth]{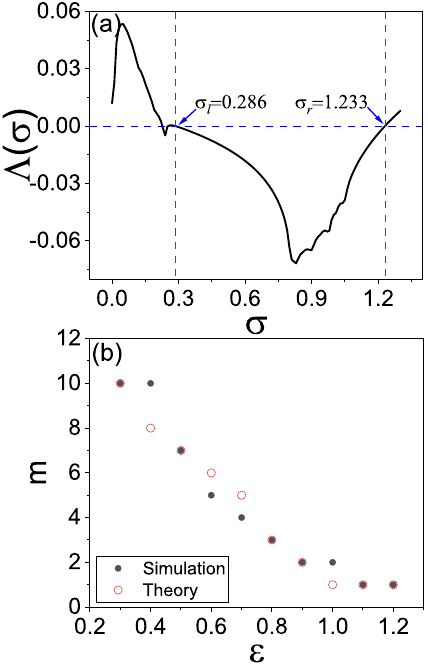}
\caption{Scalable synchronization in the network of $N=10$ identical chaotic HR oscillators. (a) The MSF curve. (b) The variation of the size of the synchronization cluster, $m$, with respect to the coupling strength, $\varepsilon$. Black dots are results obtained by numerical simulations. Red circles are results predicted by the theory.}
\vspace{-0.6cm}
\label{hr}
\end{center}
\end{figure}

For the reason that the chaotic R\"{o}ssler oscillators are unstable under very strong couplings (the trajectory diverges to infinity), we demonstrate the above desynchronization scenario in a small-size network consisting of $N=10$ chaotic HR oscillators. The dynamics of the $i$th HR oscillator in the network is governed by the equations~\cite{HR}
\begin{equation}
\label{HR}
\begin{cases}
\dot{x_{i}}=y_{i}+3x^2-x^3-z+I+\varepsilon\sum w_{ij}y_{j}, \\
\dot{y_{i}}=1-5x^2-y,\\
\dot{z_{i}}=-rz+r(4x+6.4).
\end{cases}
\end{equation}  
The parameters are chosen as $I=3.2$ and $r=6\times 10^{-3}$, by which the oscillators present chaotic motions in the isolated form. Plotted in Fig.~\ref{hr}(a) is the MSF curve obtained by simulations, which shows that $\Lambda<0$ when $\sigma\in (0.286,1.233)$. For this small-size network, we have $\lambda_N=-7.2$ and $\lambda_2=-1.2$, which gives the critical couplings $\varepsilon_N=-\sigma_r/\lambda_N\approx 0.171$ (when mode $N$ becomes unstable) and $\varepsilon_2-\sigma_r/\lambda_2\approx 1.03$ (when mode $2$ becomes unstable). Shown in Fig.~\ref{hr}(b) is the variation of the size of the synchronization cluster with respect to the coupling strength, in which the results predicted by the theory are represented by red circles and the results obtained by simulations are represented by black dots. Still, two oscillators are regarded as synchronized if after the transient  ($T=200$) the synchronization error between them is less than the threshold $\Delta x_c=0.1$. We see that the theoretical results are in good agreement with the numerical ones, both showing that the oscillators are desynchronized in sequence with the increase of the coupling strength. 

\section{Discussions and conclusion}

Compared to the CS phenomena reported in the literature, the phenomenon of scalable synchronization cluster reported in our present work possesses some distinct features, including: (1) the coexistence of a single synchronization cluster and many desynchronized oscillators; (2) the size of the synchronization cluster can be adjusted freely by tunning a single system parameter, i.e., the uniform coupling strength; (3) with the increase of the coupling strength, the desynchronized oscillators join the cluster one-by-one, making the size of the synchronization cluster increased gradually; (4) in the presence of perturbations, oscillators inside the cluster are restored to the synchronization state in sequence. This new CS phenomenon extends our knowledge on the collective behaviors of coupled chaotic oscillators, and the unique features of scalable synchronization clusters might have implications to the design and management of some real-world systems, with the details the following. 

Scalable synchronization clusters reveal a new path to synchronization in complex systems of coupled oscillators. Besides the critical point characterizing the onset of synchronization (mainly concerned for coupled non-identical phase oscillators) and the critical point for global synchronization (mainly concerned for coupled identical chaotic oscillators), attention has been also paid to the transition from desynchronization to global synchronization in coupled oscillators.~\cite{Zheng1998,SynTransition,ExplosiveSyn,AdpNet:Scholl2021,Tran:ZCS2006,Tran:HL2006,SynTrans:Fu2,CS:transition} In exploring the synchronization transition, two important issues are (1) how to identify the partially coherent states appeared at the intermediate coupling strengths and (2) how the system dynamics is transmitted among the partially coherent states as the coupling strength varies. Depending on the nodal dynamics and the network structure, the transition might present very different forms. For instance, the order parameter characterizing the synchronization degree of coupled phase oscillators might vary smoothly with the coupling strength (i.e., the type of 2nd order phase transition), or increase suddenly at some critical coupling from a small value to a very large value (i.e., the type of 1st order phase transition).~\cite{SynTransition,ExplosiveSyn} The different transition paths are attributed generally to the different synchronization scenarios of the clusters, with the former being featured by the growth of a giant synchronization cluster and the latter by the merging of several mesoscale synchronization clusters. In either scenario, the transition is non-smooth at the oscillator level, as the competition between the clusters leads to often the ``hopings" of the oscillators. This feature of non-smooth synchronization transition is more distinct for coupled chaotic oscillators, as the formation of the synchronization clusters is dependent on the mesoscale network symmetries, which are discrete and rare for complex networks.~\cite{Tran:ZCS2006,Tran:HL2006,SynTrans:Fu2,CS:transition} In some special cases, it is even shown that the increase of the coupling strength might deteriorate the synchronization degree of networked chaotic oscillators, i.e., a crossover is appeared in the transition.~\cite{Tran:HL2006} By contrast, the transition reported in our present work is smooth at the oscillator level and, more interestingly, does not present a clear transition point. Specifically, with the increase of the coupling strength, the cluster is expanded gradually by recruiting oscillators from the desynchronization background. Once synchronized, the oscillators will be staying always in the cluster, till the global synchronization is achieved. By the method of eigenvector-based analysis, we can predict not only the critical coupling where a specific oscillator is synchronized to the cluster, but also the contents of the cluster at any specific coupling, rendering the transition completely predictable and controllable. Another intriguing feature of the new transition is that, unlike the typical transitions where critical points can be defined (e.g., the 1st and 2nd order synchronization transitions), there is no critical point characterizing a distinct change of the system dynamics. This is reminiscent of the phase transition of the glassic state in liquids,~\cite{Glass:1949,Glass:parisi}, in which a linear increase of the system order parameter with the temperature is observed. (We note that the connection to the glass transition is phenomenological, as the mechanisms of the two transitions are essentially different.)     

Our studies shed new light on the relationship between CS behaviors and network symmetries. In exploring the CS behaviors of networked systems, an important finding achieved in recent years is the crucial role that network symmetries played in generating the CS states. Strictly speaking, for a complex network of identical chaotic oscillators, each symmetry of the network structure corresponds to a specific CS state, though the CS state might be unstable. For this reason, considerable attention has been given to the calculation of network symmetries in complex systems.~\cite{SCS:Pecora,CSO:Khanra}  For the network model proposed in our current study, as each node has a unique coupling capacity, the network is completely asymmetric. Counterintuitively, it is shown that this artificially synthesized asymmetric network can generate rich CS phenomena, namely a scalable synchronization cluster. The seemly abnormal CS phenomenon can be interpreted from the point of view of noise-induced synchronization, as follows. When a synchronization cluster is formed, the interactions between oscillators inside the cluster are functionally vanished (due to the linear feedback couplings). That is, oscillators inside the cluster are decoupled from each other but are all driven by the same input signals from the desynchronized oscillators. According to the mechanism of noise-induced synchronization,~\cite{NISYN:AM1994,NISYN:ZCS2002} strong coherence can emerge between uncoupled oscillators when the amplitude of the common driving signals is large enough. The abnormal CS phenomenon can also be explained from the viewpoint of EEP.~\cite{EEP:Cardoso,EEP:OClery,EEP:Schaub2016} In this picture, each desynchronized oscillator can be treated as a trivial cluster, and the number of connections between each trivial cluster and the synchronization cluster is independent of the oscillator index.    

Scalable synchronization clusters might provide an alternative approach for protecting complex networks from cascading failures. Besides synchronization, another form of collective behavior observed in a wide range of real-world systems is cascading failures.~\cite{Cas:Watts2002,Cas:Motter2002,Cas:WXG2009} Taking the power-grid network as an example, when a power station (line) is overloaded and becomes dysfunctional, the redistribution of the loads over the remaining stations (lines) might trigger a cascade of subsequent failures, resulting in a large-scale blackout such as the August 14, 2003 event occurred in the northeastern United States.~\cite{Cas:YYScience} A similar phenomenon is also observed in telecommunication and transportation networks, where the dysfunction of a few links might lead to large-scale congestion through a cascading process. For the catastrophic damages induced by cascading failures, methods have been proposed in the literature on how to prevent complex networks from cascades, e.g., updating the protecting strategies or modifying the network structures.~\cite{Cas:Motter2004,Cas:WXG2009} Our studies show that, due to the unique dynamical features of scalable synchronization cluster, the impacts of the cascading dynamics can be effectively mitigated in the proposed network model. The protection is manifested in two aspects. First, the synchronization of the oscillators inside the cluster is not affected by perturbations added to the desynchronized nodes. This feature can be seen from Eqs.~(\ref{vk}) and (\ref{transient-2}) in our analysis, which shows that perturbations added on the desynchronized oscillators (e.g., changing the oscillators states or modifying the couplings between them) contribute to only the synchronization background [the term $c(t)$ in the equation] but not the cluster-based synchronization errors. Second, if the perturbations are added on oscillators inside the cluster, Eqs.~(\ref{transient-2}) and (\ref{transient-3}) show that synchronization is restored quickly among the core oscillators (oscillators of smaller index) after a short transient. This feature is very favorable in circumstances where the basic system functions are required to be resummed first among some key elements after the perturbation, e.g., the recovery of the functionality of telecommunication systems from geomagnetic storms. 

As a final remark, we would like to note that the method of eigenvector-based analysis described in Sec. IV applies to only situations when the statistical properties of the coupled oscillators are close to that of the isolated oscillator. More specifically, in obtaining Eq.~(\ref{vk}) (which defines the contents of the clusters) and Eq.~(\ref{range}) (which predicts the critical couplings for generating the CS states), we have approximated the manifolds of the coupled oscillators by the manifold of the global synchronization state [i.e., $\mathbf{s}(t)$ in Eq.~(\ref{msf})], in evaluating the stability of the CS states. Whereas this approximation is generally valid for weak couplings (e.g., the fully desynchronized states) and strong couplings (i.e., in the vicinity of the global synchronization state), it might fail for the intermediate couplings where the oscillator attractors could be seriously deformed. In this case, to find the CS states in the transition regime, we have to rely on numerical simulations. For instance, when adopting the chaotic Logistic maps as the nodal dynamics, many periodic windows are observed in the transition regime, which can not be predicted by the proposed theory. Nevertheless, given the attractors of the oscillators are not sensitive to the coupling strength (e.g., the chaotic Lorenz and R\"{o}ssler oscillators), the method of eigenvector-based analysis is effective.

To summarize, we have investigated the CS behaviors in a synthetic network of identical chaotic oscillators, and found that the size of the synchronization cluster can be adjusted freely by changing the coupling parameter, namely the phenomenon of scalable synchronization cluster. The mechanism of scalable synchronization cluster has been investigated by the method of eigenvector-based analysis, and it is revealed that the scalability of the cluster is attributed to the hierarchical structures of the eigenvector elements. By this theoretical method, we are able to predict not only the critical coupling for generating a cluster of specific size, but also the contents of the cluster and the sequence of the oscillators inside the cluster in restoring to the synchronization state. The findings extend our knowledge on the synchronization behaviors of networked oscillators, and might be helpful for the management and design of real-world complex systems.
\vspace{-0.5cm}

\section*{Acknowledgements}

This work was supported by the National Natural Science Foundation of China (NNSFC) under Grant Nos.~12105165, 12275165, and 12005006. X.G.W. was also supported by the Fundamental Research Funds for the Central Universities under Grant No. GK202202003. H.B.Q was supported by the Natural Science Foundation of Shaanxi Province Grants No. 2022JM-004.
\vspace{-0.5cm}

\section*{Authors DECLARATIONS}

\subsection*{Conflict of Interest}
The authors have no conflicts of interest to disclose.
\vspace{-0.5cm}

\section*{Data Availability Statement}
The source codes and data that support the findings of this study are available from the corresponding author upon reasonable request.

\end{document}